\documentclass[11pt]{article}

\usepackage[normalem]{ulem}
\usepackage[swedish,english]{babel}

\usepackage{color}
\usepackage[dvipsnames]{xcolor}
\usepackage{amssymb}
\usepackage{amsmath}
\usepackage{ae}
\usepackage{slashed}
\usepackage{graphics}
\usepackage{graphicx}
\usepackage{epstopdf}
\usepackage{url}
\usepackage[linktocpage]{hyperref}

\usepackage{cite}

\usepackage{times}
\usepackage{xfrac}

\usepackage{fancyhdr}
\usepackage{enumerate}

\def\be{\begin{equation}}
\def\ee{\end{equation}}

\numberwithin{equation}{section}
\usepackage{tikz}
\begin{document}
\selectlanguage{english}
\frenchspacing
\pagenumbering{roman}
\begin{center}
\null\vspace{\stretch{1}}

{ \Large {\bf 
A paradox regarding monogamy of entanglement
}}
\\
\vspace{1cm}
Anna Karlsson$^{1,2}$
\vspace{1cm}

{\small $^{1}${\it
Institute for Advanced Study, School of Natural Sciences\\
1 Einstein Drive, Princeton, NJ 08540, USA}}
\vspace{0.5cm}

{\small $^{2}${\it
Division of Theoretical Physics, Department of Physics, \\
Chalmers University of Technology, 412 96 Gothenburg, Sweden}}
\vspace{1.6cm}
\end{center}

\begin{abstract}
In density matrix theory, entanglement is monogamous. However, we show that qubits can be arbitrarily entangled in a different, recently constructed model of qubit entanglement \cite{Karlsson:2019vpr}. We illustrate the differences between these two models, analyse how the density matrix property of monogamy of entanglement originates in assumptions of classical correlations in the construction of that model, and explain the counterexample to monogamy in the alternative model. We conclude that monogamy of entanglement is a theoretical assumption, not necessarily a physical property, and discuss how contemporary theory relies on that assumption. The properties of entanglement entropy are very different in the two models --- a priori, the entropy in the alternative model is classical.
\end{abstract}

\vspace{\stretch{3}}
\thispagestyle{empty}
\newpage
\pagenumbering{arabic}
\tableofcontents
\vspace{0.5cm}
\noindent\hrule

\section{Introduction}\label{s.Intro}
The topic of this article is how to accurately model quantum correlations. In quantum theory, quantum systems are currently modelled by density matrices $(\rho)$ and entanglement is recognized to be monogamous \cite{PhysRevA.61.052306}. Recently, a different way of modelling qubit systems was suggested in \cite{Karlsson:2019vpr}. Both models describe qubit entanglement, i.e. they capture the qubit single particle and pair correlation behaviour (the presently observed characteristics). Nevertheless, they are different correlation models, with different methods for how to detail specifications of and correlations between complementary observables. Interestingly, as we show in \S\ref{s.counterex}, they give significantly different physical predictions: they do not agree on monogamy of entanglement\footnote{Monogamy violations have previously been discussed e.g. in relation to entanglement measures \cite{PhysRevLett.117.060501} and for qubits in chronology-violating regions \cite{Marletto:2019}. Note that our discussion on entanglement is at the level of how complementary correlations are defined, and decoupled from measure and chronology aspects, as well as properties of mixed entangled states \cite{Bennett:1996gf}.}. Put simply, monogamy is not forced upon a model of entanglement by the defining properties of $\leq2$ qubits --- that requires further restrictions.

A situation where entanglement is not theoretically restricted to be monogamous, but has been assumed to be so, creates a paradox regarding what the physical properties of an accurate quantum theory really are. A counterexample to that $\rho$ by necessity gives a complete, general description of entanglement can be identified by that it is straightforward to construct three subsystems that are mutually fully entangled\footnote{We use `full entanglement' to describe correlations in the model of \cite{Karlsson:2019vpr} with one-to-one correlations at qubit measurements in the same direction (corresponding to pure ensembles in density matrix theory).
} in the alternative model of \cite{Karlsson:2019vpr}. By comparison, while $\rho$ encodes entanglement, the subsystem correlations are governed by a partial trace/purification of $\rho$,
\be\label{eq.rhoclass}
\rho_A=\operatorname{tr}_B(\rho_{AB})=\sum_\lambda\rho_A(\lambda)\sum_b P(b|\lambda)\,,\quad \begin{array}{l}\lambda:\,\text{a set of local variables}\\b:\, \text{outcomes in $B$}\end{array}
\ee
so that the correlations between separate subsystems (here $A$ and $B$) are mediated by classical correlations (within probability theory) rather than figuring in a more general setting with entangling subsystem correlations. This illustrates, as is discussed in more detail in \S\ref{ss.analysis}, a way to understand why monogamy of entanglement is a property of density matrix theory.

Recall that Bell's theorem \cite{Bell:1964kc} states that the algebra for information theoretical quantities changes in the presence of complementarity. Classical, local\footnote{The words `local' and `causal' are sometimes used interchangeably, but in this text we distinguish between the two to have a clear way of referring to the behaviour of the correlations of the theoretical model employed (local/not) vs the behaviour of information (causal).} correlations do not capture or predict entanglement. In the same way, correlations between quantum subsystems do not by theoretical necessity need to be of the classical quality \eqref{eq.rhoclass} encoded in density matrices. To consider non-monogamous entanglement is to take one step further in considering what is allowed by complementarity, compared to the change from probability theory to density matrices.

Based on the counterexample, we draw the conclusion that it is possible that the trace and purification procedures in density matrix theory fail to be faithful to quantum physics, when complementary observables are present in the state vectors. It all depends on what the physical correlations actually are. This also goes for entanglement entropy, which is built on traces of density matrices, such as the von Neumann entropy
\be
S_{vN}=-\operatorname{tr}(\rho\ln\rho)\,,
\ee
and creates a paradox in that quantum (information) theory currently build on these exact entities, and have been assumed accurate with respect to complementarity, especially in analyses of entanglement and spacetime physics. Of course, it may well be that $\rho$ captures the correct physics, but for this to be certain, the reasons for it need to be identified.

\subsection{Indications of a presence of general entanglement}\label{s.ind}
A reason to consider entanglement beyond the monogamous case can be found in the `Page curve' \cite{Page:1993wv,Page:2013dx} behaviour of the entropy of Hawking radiation. For this to be present in the gravity theory, additional non-local spacetime identifications (wormholes) must be introduced into the theory as in \cite{Penington:2019npb,Almheiri:2019psf,Almheiri:2019hni}, where additional geometric connections between the radiation and the black hole interior are introduced, while the theory remains causal.

Here, it is relevant to note a parallel. To introduce entanglement into a classical theory while keeping a classical point of view, it is necessary to make non-local identifications by hand, i.e. to introduce correlations that are not present within the local, classical theory. This can e.g. be done through extra connections between points in spacetime, as suggested in the ER=EPR conjecture \cite{Einstein:1935rr,Einstein:1935tc,Maldacena:2001kr,Swingle:2009bg,VanRaamsdonk:2010pw,Hartman:2013qma,Maldacena:2013xja}, where the classically non-local correlations given by entanglement are interpreted as wormholes\cite{Maldacena:2013xja}. A classical theory with wormholes (with two exits) effectively corresponds to a theory of monogamous entanglement without wormholes.

Although the methods in \cite{Penington:2019npb,Almheiri:2019psf,Almheiri:2019hni} only concern monogamous entanglement, the process of adding non-local spacetime identifications to a theory (including monogamous entanglement in this case) parallels the effect of mimicking entanglement through including wormholes in the classical theory. With wormholes added to a theory already including entanglement, the ensuing theory (provided that it is causal) reasonably corresponds to an effective theory without wormholes that includes entanglement beyond the monogamous restriction. Both by that the existing entanglement is extended, and by that wormholes with more than two exists are considered \cite{Akers:2019nfi}, effectively identifying more than two points in spacetime. Regardless of the initial theory (without wormholes), adding wormholes while keeping causality is one way to model extra entanglement by geometry, compared to what is inherent to the theory to begin with, and effectively has a counterpart in an extended theory without wormholes. In this sense one ends up with an effective gravity theory with more general entanglement than in the monogamous case, even if one begins with an assumption of monogamy. Some unspecified type of tripartite entanglement was also argued to be necessary in \cite{Akers:2019nfy}, but without any reference to issues with monogamy.

If one wishes to describe the extended correlations required to capture the Page curve in a gravity theory without added geometrical identifications, the entanglement model in \S\ref{s.counterex} is a candidate theory, encompassing entanglement beyond the monogamous case modelled by density matrices. One can also consider various scenarios for when general entanglement is present. For example, monogamous entanglement may be a good approximation in the semiclassical limit (without quantum gravity effects).

\subsection{Non-signalling and detection of general entanglement}
A concern when considering correlations beyond monogamous entanglement, apart from what density matrix theory proves, is whether the alteration is compatible with causality, or non-signalling. The concern extends beyond the properties of the model (non-classically local, as outlined in appendix \ref{s.details}) to the classical perception of the state configuration. For qubits, a violation of Bell's inequality \cite{Bell:1964kc} gives that no probability function can specify the qubit outcomes in three directions simultaneously; a $P(S_A,S_B,S_C|{\bf a,b,c})$ compatible with the pair correlations $P(S_A,S_B|{\bf a,b})$ does not exist\footnote{Here we refer to qubit outcomes $S$ at different sites $(A,B,C)$ given the measurement directions ${\bf a,b,c}\in S^{d-1}$.}. However, there is a loophole to this classical restriction: qubit systems from which it is impossible to extract a qubit (for measurement) while keeping track of its orientation. Then $P(S_A,S_B,S_C)=1/8$. By comparison, such a loophole makes room for complementarity to begin with: classical measurements commute, but since some measurements only can be made one at a time, non-commutativity is allowed. While such loopholes are classically counterintuitive, they are physically relevant and open up for physical behaviour incompatible with standard, classical notions. Candidate systems include correlations at the Planck scale (where the classical meaning of e.g. directions dissolves), at black hole horizons, and spin $\sfrac{1}{2}$ systems where extraction means a loss of monitored relative orientation.

Consequently, non-monogamous correlations would likely only be classically observable through the degrees of freedom of a strongly correlated system. The physical indicator would be the entropy, not statistics from outcomes of qubit measurements. We discuss the entropy in \S\ref{s.impl} and appendix \ref{s.Sco}.

\subsection{Summary and overview}
We show that the unconventional way of modelling qubit correlations of \cite{Karlsson:2019vpr} encompasses non-monogamous entanglement, contrary to what is true for density matrices, through presenting the counterexample of \eqref{eq.3co} in \S\ref{s.counterex}. Consequently, monogamy of entanglement requires more support from physical observations than presently identified. Its validity for qubits cannot be proven only based on observations of the single particle behaviour and the pair correlations, since two different models with opposite results can be formed in that scenario. Whether or not entanglement between several subsystems close to \eqref{eq.3co} is physically attainable is an open question.

Density matrix theory risks failing to capture physical properties since restrictions are imposed on the subsystem correlations by the partial trace and purification procedures. By construction, the model does not encode entanglement beyond the monogamous case. We analyse how the density matrix restriction can be understood in \S\ref{ss.analysis}. Our consideration is motivated by that classical relations, including the density matrix subsystem correlations of \eqref{eq.rhoclass}, should not be assumed to be accurate without thorough justification in the presence of complementarity, i.e. for correlations between complementary observables. Further details on the comparison between the classical theory, density matrix theory, and the model of \cite{Karlsson:2019vpr} can be found in appendix \ref{s.details}. Since density matrix subsystem correlations are mediated by classical correlations, we conclude that monogamy of entanglement is an assumption that goes into the formalism, not a result from it.

Following the counterexample, the question is what the proper model of entanglement is, and to what extent previous results on entanglement are accurate. We discuss implications for how entangled systems can be modelled in \S\ref{s.impl} and introduce a preliminary concept of entropy $S_{oi}$, formed (as detailed in appendix \ref{s.Sco}) through an imitation of how Gibbs entropy relies on degrees of freedom of microstates and by taking unique states in the model in \cite{Karlsson:2019vpr} to define configurations of the qubit system. This entropy is not guaranteed to be physically accurate, and it might be subject to further restrictions, but the construction does not include additional assumptions compared to the construction of entanglement entropy. That is, the degrees of freedom of a given model are taken at face value. Curiously, its behaviour is classical, without the characteristic features associated with entanglement entropy, indicating that more consideration may need to go into formulations of entropy in general. At the end we discuss general implications for the model discrepancy, including that the alternative entropy of a subsystem a priori would seem proportionate to the subsystem volume instead of the boundary area.

It is evident that if restrictions are imposed in quantum models without direct motivation from physical observations, there is a risk of losing physical properties in the model. Especial care should be taken with respect to classical assumptions when entanglement and properties of complementary observables are analysed. A critical question for entanglement now is: what are the reasons for monogamy of entanglement apart from the defining properties of density matrix theory?

\section{Analysis of the partial trace}\label{ss.analysis}
So far, monogamy of entanglement has been treated as a general result for entanglement, a condition set by density matrix theory. We will elaborate on the counterexample of \eqref{eq.3co} shortly, but first we analyse how correlations are encoded in density matrices.

In density matrix theory, the classical theory is extended by that the probability distribution is replaced by
\be\label{eq.rho}
\rho=\sum_{i=1}^np_i|\psi_i\rangle\langle\psi_i|\,,\quad |\psi_i\rangle=\sum_{jk}q_{ijk} |a_{j}\rangle_A \otimes |b_{k}\rangle_B\otimes \ldots
\ee
for some number of Hilbert spaces ($A,B$ etc.). Here, $p_i$ represents probabilities for different states $\psi_i$, and $a_{j}, b_{k}$ represent outcomes forming orthonormal bases for the respective Hilbert spaces. By that subdivisions of quantum systems are captured through tracing out parts of $\rho$ (and correspondingly for the reverse process, purification) a general, pure $\rho_{AB}$ (with $A$ equivalently substituting for some larger Hilbert space) is reduced through
\be
\rho_A=\operatorname{tr}_B(\rho_{AB})=\sum_{jkj'k'}\langle b_{k'}|b_k\rangle q_{jk}q_{j'k'}|a_j\rangle\langle a_{j'}| =\sum_\lambda\rho_A(\lambda) \sum_b P(b|\lambda)\,.
\ee
The trace over $B$ gives rise to $\langle b_{k'}|b_k\rangle$ and describes correlations equivalently mediated through $P(b|\lambda)$, for some set of variables $\lambda$. The division is different from the classical, local correlation
\be
\sum_{b} P(a,b|\lambda)= P(a|\lambda) \sum_{b} P(b|\lambda)
\ee
in that the reduction is to $\rho_A$, but the correlations between the subsystems $A$ and $B$ described by the partial trace, $\operatorname{tr}_B(\rho_{AB})$, are mediated through a classical entity.

This illustrates how the correlations taken into account in a subdivision of $\rho$ are mediated by classical correlations, and provides a way to understand why entanglement is monogamous in the formalism. Even though $\rho$ is an extension of probability theory and encodes more extensive correlations by its definition \eqref{eq.rho}, in subdivisions of/additions to $\rho$ the processes of trace/purification infer that the only additional correlations accommodated are mediated by the classical theory. Note that this constitutes an assumption that goes into the formalism, and does not represent a prediction on the behaviour of entanglement in general. It is a property that defines how far the model reaches, unconnected to the model compatibility of the trace/purification procedures. By construction, the only entanglement described is monogamous.

\section{A counterexample to monogamy of entanglement}\label{s.counterex}
For the counterexample, we turn to the alternative qubit correlation model in \cite{Karlsson:2019vpr}. Since it is based on that the Bell correlations can be captured by a model where complementary observables are not simultaneously quantified (even theoretically, in contrast to in $\rho$), through a presence of information that is orthogonal to classical information, we call it an `orthogonal information model'. Of the different ways that exist for modelling correlations that are not local and classical, mainly dividable into `non-local identifications' (e.g. wormholes and $\rho$) and `non-classical information' (e.g. negative probabilities and $\rho$), the orthogonal information model falls into the second category.

In \cite{Karlsson:2019vpr}, the qubit correlations (we use spin $\sfrac{1}{2}$ for the example) are captured by that the observables are obtained from a projection of an extended information entity $G\in\mathbb{H}$. For a single particle,
\be
\langle S_A({\bf a})\rangle=\operatorname{Re} [ G({\bf a}|\sigma_A)]=s\,{\bf a \cdot r}\,,
\ee
with directions ${\bf a,r}\in S^2$ (unit 2-sphere), spin outcomes $S_A\in\{\pm1\}$ and $\sigma$ denoting a set of variables specifying the orientation of a coordinate system (where ${\bf r}=\hat x$):
\begin{subequations}\label{eq.G}
\be
\sigma=\{{\bf r},\hat y,\hat z, s\}\,,\quad G({\bf a}|\sigma_A)=se^{\theta_{ar}{\bf u}}=s(\cos\theta_{ar}+\sin\theta_{ar}(u_yj+u_zk))\,, 
\ee
\be
s\in\{\pm1\}\,,\quad\cos\theta_{ar}={\bf a\cdot r}\,,\quad {\bf u}={\bf r\times a}/|{\bf r\times a}|\,,\quad i^2=j^2=k^2=ijk=-1\,,
\ee
\end{subequations}
where $|{\bf r\times a}|=\sin\theta_{ar}$, and ${\bf u}={\bf 0}$ if ${\bf a\parallel r}$. Due to the two-level outcomes of $S$, the probabilities can be retrieved from the expectation values through $2P(\cdot)-1=\langle\cdot\rangle$. With $G^*$ representing the conjugate of $G$, correlations between pairs of qubits are captured by
\be\label{eq.pair}
\langle S_A({\bf a})S_B({\bf b})\rangle=\operatorname{Re}[ G^*({\bf b}|\sigma_B)G({\bf a}|\sigma_A)]\,,
\ee
which gives a full overlap for $\sigma_A=\sigma_B$ when ${\bf a}={\bf b}$ (e.g. one and the same information entity).The Bell correlations are captured by that pair production is characterized by
\be
\sigma_B=\sigma_A\big|_{s\rightarrow-s}\,:\quad \eqref{eq.pair}=-{\bf a\cdot b}\,.
\ee

This is \emph{a model} of the Bell correlations, very different from $\rho$ except in that the concept of information has been extended to include parts without a classical interpretation. Instead of that complementary correlations are encoded in off-diagonal elements of $\rho$, they appear in parts $\not\in\mathbb{R}$ that can combine to give real contributions to the observable expectation values. What is returned from a measurement depends on what is probed, through $G({\bf a}|\sigma)$. 

In the orthogonal information model, the non-classicality lies in the structure $\in\mathbb{H}$, not in the state specification given by $\sigma$. Each $\sigma$ defines the properties of \emph{one particle}, and summations over different particles is a classical operation. (Compare with a sum over different outcome configurations for one and the same particle, such as outcomes $\pm1$ in $\rho$, which is not decoupled from complementarity.) Because $\sigma$ defines single information entities, and the structure $\in \mathbb{H}$ takes care of the correlations, mutual entangling correlations between many particles are readily available through
\begin{subequations}\label{eq.3co}
\be
\sigma_A\equiv\sigma_B\equiv\sigma_C\,:
\ee
\be
\langle S_i({\bf a})S_j({\bf b})\rangle=\operatorname{Re}[ G^*({\bf b}|\sigma_j)G({\bf a}|\sigma_i)]={\bf a \cdot b}\,,\quad \forall i\,,j\in\{A,B,C\}\,.
\ee
\end{subequations}
This illustrates the counterexample: a model of the qubit correlations where, theoretically, any number of qubits can be fully entangled\footnote{Note that full entanglement between multiple quantum states is compatible with the no-cloning theorem \cite{Wootters:1982zz}, which is a statement on the impossibility of producing a clone of a given state, not about the production of multiple copies of a random state. An example of the latter is the linear polarization of pair produced photons.}.

\section{Implications for entangled systems}\label{s.impl}
The possibly restrictive nature of the trace/purification operations creates a problem regarding to what degree current quantum (information) theory results are accurate. Any quantization not coupled to complementarity is fine, since those represent discretizations where the physics is described by classical correlations. In the presence of complementarity, a first question must concern the accuracy of the suggested model in \cite{Karlsson:2019vpr}. Since it is a valid model of the currently observed qubit behaviour (single particle behaviour and pair correlations), however unconventional the construction is, further physical properties are required to distinguish between the two models in terms of physical accuracy. A second key question is in what ways the two models give rise to different physical results, apart from or as a result from (non-)monogamy of entanglement.

Here, entropy is useful for comparing physical results and general properties of the models. It is straightforward to base a concept of qubit entropy directly on the degrees of freedom of the individual state in the orthogonal information model, described by $\sigma$, assuming \emph{(i)} that the entropy can be based on the state degrees of freedom in the same way as for Gibbs entropy, despite the presence of complementarity, and \emph{(ii)} that no further, not yet identified, restrictions apply. The construction can be found in appendix \ref{s.Sco}. Since $\sigma$ behaves classically, the subsequent qubit entropy has a classical behaviour, contrary to the behaviour of entanglement entropy. This might be an indication of that further restrictions should be expected.

The entropy obtained in this way (as described in appendix \ref{s.Sco}) is preliminary, but represents a best first ansatz and readily illustrates a disagreement with the result of monogamy of entanglement. The properties of \eqref{eq.3co} can equally be given in terms of entropy (here with $S=S_{oi}$) as
\begin{subequations}\label{eq.Soiex}
\be
S({ABC})=S({AB})=S({AC})=S({BC})=S(X)=S_{oi,\,\text{qubit}}\,,\quad X\in\{A,B,C\}\,:
\ee
\be
S(A|B)=S(A|C)=S(B|C)=S(AB|C)=S(AC|B)=S(BC|A)=0\,,
\ee
\end{subequations}
since there is only one set of qubit degrees of freedom. The full system is set in direct relation to any subsystem configuration. Note the difference in the minimal entropy described (a pure state has $S_{oi}>0$) compared with entanglement entropy, and that the conditional entropy is non-negative. While the counterexample of \eqref{eq.3co} suffices to illustrate non-monogamy of the related entanglement, the entropy provides a more versatile way of characterizing the states and their intercorrelations.

At least one of the two entropy models must be subject to further constraints or freedoms, due to the physical disagreement on monogamy of entanglement. Still, many entropy properties can be expected to hold for both models. Since the entropy based on the orthogonal information model has a classical behaviour, and (naively) all the properties that goes with it, important shared qualities include positivity of entropy, subadditivity and strong subadditivity. The real difference would seem limited to consequences of purification and traces over subsystems.

An interesting aspect is that while the ortogonal information model straightforwardly can reduce to a monogamous setting through restrictions on $\sigma$, the same is not true for a reduction of $S_{oi}$ to entanglement entropy (including negative conditional entropy). For this, the notion of entropy would have to be connected to the degrees of freedom of $\rho$, specifically, for some reason. For example, a pure Bell state would have to be identified to have one degree of freedom only, instead of the same number of degrees of freedom as a single qubit, as is the case in \eqref{eq.Soiex}.

In its present form, the classical behaviour of the entropy $S_{oi}$ brings about an important complication. A very useful feature of entanglement entropy is reliant on the partial trace: its proportionality to area instead of volume \cite{Bombelli:1986rw,Srednicki:1993im}, in the sense that the entanglement entropy of a subsystem of a pure state is dependent on the subsystem boundary. This gives it a decided similarity to the Bekenstein--Hawking entropy \cite{Bekenstein:1973ur,Hawking:1976de} of black holes. It is difficult to see how this proportionality would arise in general without the partial trace, equivalently the relation \eqref{eq.rhoclass}. The preliminary entropy $S_{oi}$ would be proportional to volume, except for very special entanglement configurations.

In short, the physics seems to allow for general entanglement (as in the orthogonal information model). If one assumes monogamy of entanglement, the behaviour of the Hawking radiation indicates a presence of more general entanglement, e.g. through `wormhole' identifications capturing the Page curve. If one instead assumes general entanglement to be present, the associated entropy appears proportionate to subsystem volume, without a direct connection to the behaviour of the Bekenstein--Hawking entropy, or the Ryu--Takayanagi formula \cite{Ryu:2006bv,Hubeny:2007xt}. How to accurately model black holes without monogamy of entanglement then would seem an open question.

We do not address the physical existence of non-monogamous entanglement here, only the theoretical possibility. The present critique is that it is premature to claim monogamy of entanglement in general based on that same property of density matrices, and on the currently available experimental observations. Granted, more general entanglement has not been observed, but that in itself in not conclusive proof of its absence.

The possibility of modelling entanglement beyond the monogamous case shows that it is relevant to scrutinize the assumptions made in modelling entanglement. The question of how to model physics in the presence of complementarity has a fundamental paradox in its non-classicality, and the correlations encoded are highly relevant: they impact the final result, e.g. monogamy of entanglement. Classical correlations such as by tracing out degrees of freedom (corresponding to smearing in quantum field theory) amounts to restraining the physics, which must be motivated beyond internal model consistency. For a complete, physical picture of quantum entanglement, it is crucial to understand the effects of complementarity.

\section*{Acknowledgements}
This work is supported by the Swedish Research Council grant 2017-00328.

\appendix

\section{More details on the different correlation models}\label{s.details}
In classical theory, the probability of any two outcomes ($S_A,S_B$) of two measurements (here along ${\bf a},{\bf b}\in S^{d-1}$, $d\in\{2,3\}$) in local, separate subsystems $(A,B)$ is described by
\begin{subequations}\label{e.class}
\begin{align}
&P({\bf a},S_A;{\bf b},S_B)=\int d\lambda\,\rho(\lambda)\,P({\bf a},S_A;{\bf b},S_B|\lambda)\,,\\ &P({\bf a},S_A;{\bf b},S_B|\lambda)=P({\bf a},S_A|\lambda)P({\bf b},S_B|\lambda)\,,
\end{align}
\end{subequations}
where $\lambda$ is a set of local variables and $\rho(\lambda)$ is the probability density function thereof. Bell's theorem \cite{Bell:1964kc} states that two-level systems only can be modelled in this way if the correlations satisfy Bell's inequality (listed in \eqref{eq.Belineq}). Note that while Bell's theorem often is discussed in terms of (classical) locality of correlations, it is equally a statement on \emph{the reach of the predictiveness} of the (local) classical theory. Classical theory\footnote{Excepting non-local identifications of the kind discussed in \S\ref{s.ind}.} cannot model qubit entanglement, e.g. the pair produced spin $\sfrac{1}{2}$ correlations
\be
\label{eq.spinex}
\langle S_A({\bf a})S_B({\bf b})\rangle=-{\bf a\cdot b}\,, \quad S_A,S_B\in \{\pm 1\}\,.
\ee
The correlations simply lie out of reach in terms of what \eqref{e.class} can produce.
\\\\
In density matrix theory, the classical theory is extended by that the probability distribution is replaced by \eqref{eq.rho}. A density matrix diagonal in terms of the outcomes reduces to the classical theory. The new physics lies in the off-diagonal elements of $\rho$: the extension allows for single states ($n=1$, pure states) capable of communicating one-to-one correlations (in a specific outcome basis) between two particles, despite that the observables have more than one outcome, such as for the Bell correlations. A standard example is the Bell state
\be
|\psi_{bell}\rangle=\frac{1}{\sqrt{2}}(|++\rangle+|--\rangle)\,,
\ee
with the expectation value $\langle v \rangle=\operatorname{tr}(\hat v \rho_{bell})$ for some operator $\hat v$. 

Density matrix theory successfully models pair produced entanglement, but has the interesting property of excluding entanglement shared between more than two subsystems; the entanglement is monogamous \cite{PhysRevA.61.052306}. The property is simple to illustrate for maximal entanglement. The only consistent way of obtaining the density matrix for a subsystem (i.e. to subdivide a system though reducing the number of Hilbert spaces considered) is to take a trace over the degrees of freedom that are to be disregarded. As a consequence, the ansatz one would make for a pure state with a theoretical three identical outcomes (Bell correlations between any two particles out of the three), a GHZ state \cite{GHZ},
\begin{subequations}\label{eq.+++}
\begin{align}
&|\psi_{ABC}\rangle=\frac{1}{\sqrt{2}}(|+++\rangle+|---\rangle)\,,\label{eq.3+}\\
&\operatorname{tr}_C(\rho_{ABC})=\frac{|++\rangle\langle++|+|--\rangle\langle--|}{2}\neq \rho_{bell}\,,
\end{align}
\end{subequations}
cannot reduce to a Bell state when one particle is disregarded. Instead the resulting $\rho$ is classical, incapable of encoding entanglement. A way to understand the density matrix property of monogamy of entanglement in general was discussed in \S\ref{ss.analysis}.
\\\\
In contrast to what is true for density matrices, the orthogonal information model allows for an unspecified number of states to have pair correlations that violate Bell's inequality. With suitable angles of measurement, the configuration in \eqref{eq.3co} directly corresponds to \eqref{eq.3+}, yet in \eqref{eq.3co} the three subsystems all are pairwise correlated as Bell states are. The correlations have been extended, as outlined in \S\ref{s.counterex}, and we now go into more detail about what that model entails.

\subsection*{Properties of the orthogonal information model}
The orthogonal information model constitutes a correlation model where each state is associated with a quaternion information entity $G$, and the quaternion algebra takes care of the correlations between pairs of information entities. This allows for a richer correlation model than within $\rho$. Importantly, both models capture the observed characteristics of qubit behaviour for single particles (at consecutive measurements), and from pair production. As shown in \S\ref{s.counterex}, this is not enough to distinguish between (non-)monogamy of entanglement; other physical properties need to be identified to either restrict this property of entanglement, or to verify an additional freedom which so far has been bypassed in favour of monogamy.

In specific, the orthogonal information model is constructed from that if one wants separate outcomes described by ${\bf a\cdot r}$ (expectation value overlap at consecutive measurements of single particles) to add up through a process of multiplication to ${\bf a\cdot b}$ (the pair produced overlap) while probability theory is left as an intact subset of the total correlation model, the entity considered must be extended beyond what is observable (subsequently interpreted as real projections), to a complex/quaternion model of rotation (also discussed in \cite{Karlsson:2018tod}), as in the following simple example with $\theta_{ij}=\theta_i-\theta_j\in S^1$,
\be
\text{`}\cos\theta_{ar} \times \cos\theta_{br} = \cos\theta_{ab}\text{'} \quad\text{through real projections of}\quad e^{i\theta_{ar}}e^{-i\theta_{br}}\,.
\ee
The algebra of the orthogonal entities (quaternion for spin $\sfrac{1}{2}$) then define the correlations. \cite{Karlsson:2019vpr}

That each $G(\sigma_i)$ can be treated as an individual, separate entity with general correlations to every other such entity through \eqref{eq.pair} is what allows for a consideration of more extensively entangled subsystems than in $\rho$. Where $\rho$ gives a full system specification, a quantum system specification in the orthogonal information model is built from individual qubit entities. 

In addition, each $G$ is allowed to evolve separately and locally through changes in $\sigma$. Those variables figure as a complete basis for the degrees of freedom of the qubit information entity $G$, but are decoupled from complementarity, which is taken care of by the structure $\in \mathbb{H}$. A sum over different values in $\sigma$ is a classical operation over different particles, as opposed to sums over complementary observables in specifications of $\rho$, which include different outcome scenarios for one and the same particle. \cite{Karlsson:2019vpr}

A complication is that the entries $\not\in\mathbb{R}$ lack a classical interpretation, as also is the case for the correlations. Compared to in classical theory, where all state variables reasonably should have a classical interpretation, this is allowed by a loophole furnished by complementarity: since complementary observables cannot physically be inferred to have a simultaneous classical meaning, the state specification does not need to make sense in that way either \cite{Karlsson:2019vpr}. The result is an unconventional quantum correlation model that needs to be examined further for a complete verdict on its usefulness. The discrepancy in (non-)monogamy of entanglement makes such examinations relevant.

\subsection*{Short comment on the interactions required for the correlations}
For non-monogamous entanglement to exist, suitable interactions would have to exist, together with a set of conservation laws (more general than what holds for pair creation) governing those interactions. For example, conservation of total angular momentum must be present, which for pair production gives opposite spin, but further characteristics may be relevant. A good candidate for this type of interaction would be interactions between spin $\sfrac{1}{2}$ particles, since those are known to readily interact. A question would be if the final product of such an interaction is random or dependent on the initial configurations in an equilibrating manner. Such interactions could equilibrate states towards a limit of randomly produced clones. Leaving aside the extreme example of \eqref{eq.3co}, a more reasonable conjecture is an existence of multiple states close to one another, $\sigma_i\approx\sigma_j$, where each pair randomly picked out of a (subset of a) large ensemble is entangled (compare to appendix \ref{s.toldev}) in a way not restricted by monogamy. A type of system that might be of interest for further analysis of qubit interactions is spin systems correlated through their magnetic dipole moments.

\section{Entropy in the orthogonal information model}\label{s.Sco}
\begin{figure}
\begin{center}
\includegraphics[width=5cm]{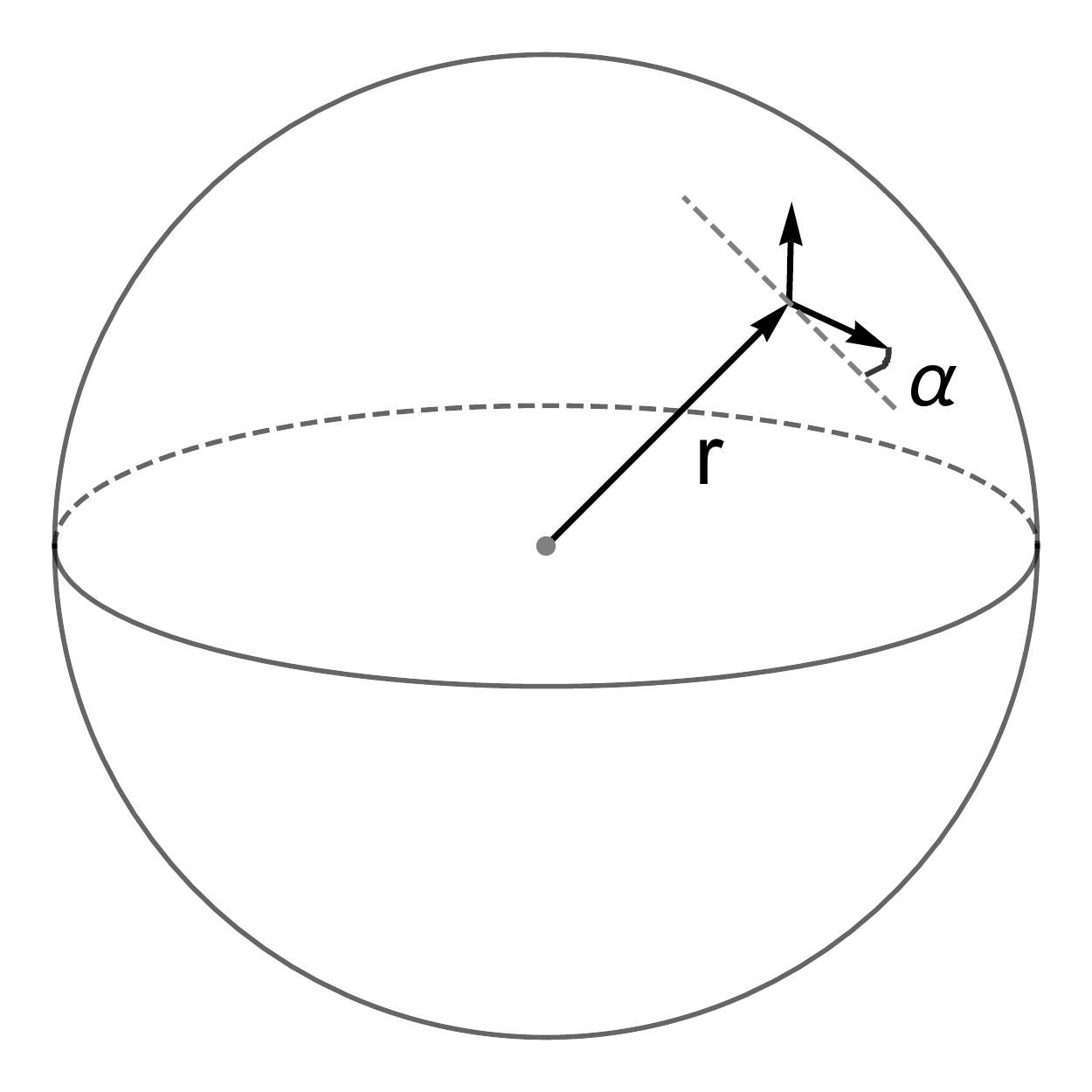}
\hspace{1cm}
\includegraphics[width=4.5cm]{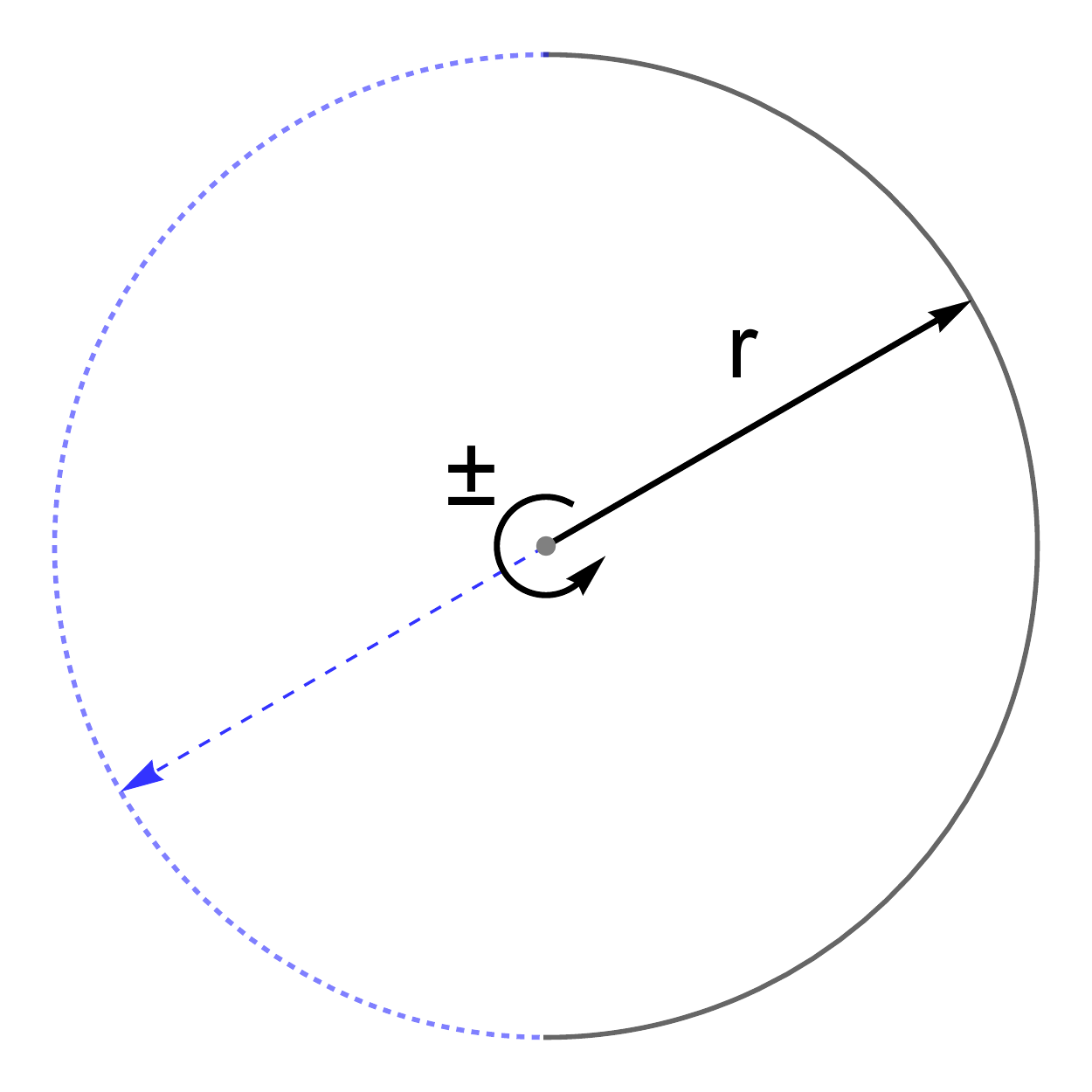}
\end{center}
\vspace{-0.5cm}
\caption{Illustration of what the qubit state specification of the orthogonal information model encodes. The variables of $\sigma$ equivalently specify the base vectors of a coordinate system in $d\in\{2,3\}$ space dimensions, through a vector ${\bf r}\in S^{d-1}$ and a rotation basis. The latter is in $2d$ \emph{(right)} given by a sign ($s$: positive/negative rotation by $s\hat p$ in the plane perpendicular to the momentum of the particle) and in $3d$ \emph{(left)} it specifies the position of the remaining two axes $(\hat y,\hat z)$ in a coordinate system with positive/negative orientation ($s$). $G({\bf a}|\sigma)$ in $3d$ is specified in \eqref{eq.G}, in $2d$ it is $G({\bf a}|{\bf r},s)=e^{2is\theta_{ar}}$ instead \cite{Karlsson:2019vpr}. In $3d$, the degrees of freedom are ${\bf r}\in S^2$, $\alpha\in[0,2\pi)$ for the position of $(\hat y,\hat z)$, and $s$. All vectors are mutually orthogonal. The $2d$ case is a subset of the $3d$ configuration, with ${\bf r}$ only covering half of $S^1$ uniquely (the model is indifferent to $\pm{\bf r}$), and no dependence on $\alpha$. \label{f.dofs}}
\end{figure}

The qubit state in the orthogonal information model is specified by $\sigma$, which describes the orientation of a two- or three-dimensional spatial system, as illustrated in figure \ref{f.dofs}. 
These quantities are not directly available through measurements, but possible to infer (as a candidate model) through the Bell pair correlations. Consequently, $\sigma$ specifies a state configuration, resembling a quantum equivalent of a microstate in statistical mechanics.

Suppose that the degrees of freedom of $\sigma$ give rise to entropy in the same way as in the classical case, as discussed in \S\ref{s.impl}, despite their non-classical origin/nature. Then, if an entropy is based on this state configuration instead of on the standard qubit two-level state vector specifying measurement outcomes, the entropy is more similar to Gibbs or Shannon entropy,
\be\label{eq.S}
S=-C \sum_i p_i \ln p_i \,, \quad C\in\{k_B,1\}\,,
\ee
than entanglement entropy, which uses density matrices instead of probabilities, such as the von Neumann entropy (directly connecting to Shannon entropy for diagonal matrices). The internal qubit degrees of freedom are readily compared to different microstates in the sense of statistical thermodynamics (although a connection through the Boltzmann constant $k_B$ is not inferred). A connection to Shannon entropy is not clear, since extractible information has no direct connection to an internal system that cannot be directly observed, but since the Gibbs and Shannon entropies overlap up to a constant which in the quantum case so far is undetermined anyway, the difference in analogy is of little relevance. The entropy in \eqref{eq.S} generalises directly to the case of orthogonal information through the probability of a state having given values for the variables that define the state, with each set of values defining a unique state,
\be\label{eq.2dDofs}
2d:\quad p=p(\theta,s)\,,\quad \theta\in [0,\pi)\,,\quad s\in\{\pm1\}\,,
\ee
\be
3d: p=p({\bf r},\alpha,s)\,,\quad {\bf r}\in S^2\,,\quad \alpha\in[0,2\pi)\,\quad s\in\{\pm1\}\,.
\ee
These degrees of freedom (compare with figure \ref{f.dofs}) begin with the direction of one of the base vectors, ${\bf r}$, which covers an angular space of $4\pi$ in $3d$ and $\pi$ in $2d$ (${\bf r}$ is replaced by $\theta$ in \eqref{eq.2dDofs} since only half of $S^1$ is covered uniquely). In addition, in $3d$ there is an angular freedom $\alpha$ of the (positive) rotation basis, which can be thought of as the relative position of a second base vector, and in both cases there is also a freedom of positive/negative orientation, $s$. For a constant probability and with the sum in \eqref{eq.S} substituted with an integral where suitable,
\be
S_{vN,\,\text{qubit}}=\ln 2 \quad\rightarrow\quad S_{oi,\,\text{qubit}}=\left\{\begin{array}{lc}c_2\times\ln 2\pi\,,&d=2\,,\\c_3\times2\ln4\pi\,,&d=3\,,\end{array}\right.
\ee
where $(c_2,c_3)$ denote constants relating the state degrees of freedom with an entropy value. For a common point of reference in the models, one can consider $c_d(d-1)\ln[(d-1)2\pi]=\ln 2$.

As for entanglement entropy, the entropy $S_{oi}$ encodes correlations in general, without being restricted to entanglement. Its `quantum' quality lies in how the degrees of freedom relate to classical results through $G$ as in \eqref{eq.G} and \eqref{eq.pair}, instead of properties that cannot be fit within a classical entropy model (compare with negative entanglement entropy). For details on what fits within the classical model and how to discern entanglement, see appendix \ref{s.toldev}.

Importantly, by describing the qubit state as in the orthogonal information model, the entropy follows classical information theory more closely than current quantum information theory. In the qubit example, the entropy is set by how states relate to each other by means of classical degrees of freedom instead of by density matrices. Since classical degrees of freedom only are reduced by a one-to-one matching between different subsystems,
\be
\operatorname{max}\left[S_{oi}(A),S_{oi}(B)\right] \leq S_{oi}({AB})\leq S_{oi}(A)+S_{oi}(B) \,,
\ee
and the conditional entropy is non-negative,
\be
S_{oi}(A|B)=S_{oi}(AB)-S_{oi}(B)\geq0\,,
\ee
same as for classical entropy, different from the entanglement entropy result. Full entanglement is represented by $S_{oi}(A|B)=0$, which for two subsystems with an equal number of degrees of freedom means $S_{oi}(AB)=S_{oi}(A)=S_{oi}(B)$. These examples constitute some first differences between the models, together with (non-)monogamy of entanglement.

Basically, the behaviour of $S_{oi}$ is classical. Consequently, it is characterised by the standard classical features of positivity of entropy, subadditivity and strong subadditivity etc. Conditional entropy, mutual information and (monotonicity of) relative entropy work the same way as in the classical theory as well. In the presence of complementarity, the features unique to quantum information theory (without special consideration of complementarity), such as negativity of conditional entropy, would appear absent.

Note that the suggested entropy in the presence of complementarity is not a complete concept:
\begin{enumerate}\itemsep0em
\item It concerns only qubits, since the behaviour of other complementary observables is not expected to follow the qubit model in \cite{Karlsson:2019vpr}.
\item Further restrictions on it cannot be ruled out. There might be conditions that cannot be observed through the pair correlations. The present entropy model simply represents the most straightforward ansatz, based on the state degrees of freedom of the orthogonal information model.
\item Complementarity is assumed not to complicate how an entropy is to be modelled. The possible issue here is how entropy is \emph{sourced}, not the correlations. The variables of $\sigma$ were inferred by the specific factorisation of information entities sought in \cite{Karlsson:2019vpr}, and are decoupled from measurement outcomes of complementary observables. These variables specify a state configuration and can be consistently summed over (as in classical probability theory) since any such sum is over different particles and therefore classical. Because of this, complementarity (which concerns correlations) should not complicate an extension to an entropy concept in a thermodynamic fashion, but the inferred state configuration is assumed to have a role analogous to that of a classical microstate. In comparison, the critique of entanglement entropy concerns the correlations, since a trace over observables is imposed without further consideration of complementarity, i.e. non-classicality of correlations between said observables, despite that the trace imposes a relation as in \eqref{eq.rhoclass}.
\item It might be more intuitive from an observer point of view to base the entropy concept on measurement outcomes instead, with some suitable modification of how subsystems are correlated, beyond a reliance on $\operatorname{tr}(\rho)$. However, the precise advantage and formulation of such a construction are unclear.
\end{enumerate}

\section{Correlations vs entanglement: a tolerance for deviations}\label{s.toldev}
It is useful to have a way to distinguish which correlations represent entanglement, especially in the orthogonal information model, where the entropy has a classical behaviour. The qubit correlations are of the form 
\be
\langle S_1({\bf a})S_2({\bf b})\rangle=y\,{\bf a \cdot b}\,,\quad y\in[-1,1]\,.
\ee
The difference between $2$ and $3d$ is only a matter of periodicity: $\cos(2\theta_{ab})$ vs ${\bf a \cdot b}=\cos\theta_{ab}$. For these correlations, Bell's inequality\footnote{Without taking experimental error (or deviations from the exactness of the quantum physical prediction) into account, as captured by the CHSH formula \cite{Clauser:1969ny}.} \cite{Bell:1964kc}
\be\label{eq.Belineq}
|\tilde P({\bf a},{\bf b})-\tilde P({\bf a,c})|\leq1-x \tilde P({\bf b,\bf c}) \,, \quad \forall {\bf a,b,c}\in S^{d-1} \,,\, \tilde P\in[-1,1]\,, \, x=\operatorname{sgn}[\tilde P({\bf b,b})],
\ee
where $\tilde P$ is a shifted probability,\footnote{The sign $x$ keeps the relation valid for both correlated and anticorrelated outcomes ($S_A,S_B$), such as for pair produced photons (linear polarization, $d=2$) and spin $\sfrac{1}{2}$ particles ($d=3$). }
\be
\tilde P({\bf a},{\bf b})=2P\big(S_A({\bf a})=S_B({\bf b})\big)-1=\langle S_A({\bf a})S_B({\bf b})\rangle\,,\quad S_A,S_B\in\{\pm1\}\,.
\ee
gives at hand that the requirement for entanglement is
\be
|y|>2/3\,.
\ee
This condition can be derived from the maximum of $|\tilde P({\bf a},{\bf b})-\tilde P({\bf a,c})|+x \tilde P({\bf b,\bf c})$. Entanglement, and hence a violation of Bell's inequality, is present for weaker correlations that $|y|=1$, and it is independent of the (anti-)correlation between the considered particles (the sign of $y$).

Entanglement in terms of a less than one-to-one correlation by an overall factor $|y|\in (2/3,1)$ is of interest e.g. in a scenario where some interaction between two qubits (other than pair production) has generated an overlap, but with an uncertainty in alignment e.g. through ${\bf r}_1\cdot {\bf r}_2\geq \cos\Delta$. We now illustrate the forms that can take.

In $2d$ the qubit state in the orthogonal information model is described by $\sigma=\{{\bf r},s\}$. Assuming that the uncertainties in the different variables are uncorrelated, and that ${\bf r}_1\cdot {\bf r}_2$ has a constant probability distribution within the considered interval, the outcomes $S_1, S_2\in \{\pm1\}$ are correlated as
\be\label{eq.2avT}
\left\langle S_1({\bf a})S_2({\bf b})\right\rangle=P(s_1=s_2) \frac{1}{2\Delta}\int_{-\Delta}^\Delta d\theta\cos(2[\theta_{ab}+\theta])=P(s_1=s_2)\frac{\sin2\Delta}{2\Delta}\cos2\theta_{ab}\,.
\ee
This is inferred by that the overlap comes from 
\be
\operatorname{Re}\left[G^*({\bf b}| {\bf r}_2,s_2)G({\bf a}| {\bf r}_1,s_1)\right]=\left[G_{2d}({\bf a}| {\bf r}_1,s_1)=e^{2is_1\theta_{ar_1}}\right]= \operatorname{Re}\left(e^{-2is_2\theta_{br_2}}e^{2is_1\theta_{ar_1}}\right)\,,
\ee
with $\theta_{ij}=\theta_i-\theta_j$, $|\theta_{ab}|$ fix, $\theta_{r_2r_1}=\theta$ and $\theta_i$ otherwise random within $I=[0,2\pi)$. For a detailed explanation, see \cite{Karlsson:2019vpr}. Contributions from $s_1=-s_2$ average to zero, since the rotation directions are mismatched. Consequently, in conditions where $s_1=s_2$ are directly correlated, and there are no preferred values for ${\bf r}_1\cdot {\bf r}_2\geq\cos\Delta$, entanglement is present for $\Delta<L/4$ in $2d$.

In $3d$ the state is described by $\sigma=\{{\bf r},\hat y, \hat z, s\}$, with ${\bf r}=\hat x$, and has a general overlap of
\be\label{eq.generalO}
\left\langle S_1({\bf a})S_2({\bf b})\right\rangle=\left\langle s_1s_2\sum_{i=1}^3 ({\bf a}\cdot \hat e_{1,i})({\bf b}\cdot \hat e_{2,i})\right\rangle\,.
\ee
This is equivalent to taking $\langle\operatorname{Re}(G^*_2G_1)\rangle$, and makes it easier to deal with the degrees of freedom inferred by different $({\bf r}, \hat y, \hat z)$ of the considered particles. In this way, the correlations between the qubits can be defined by
\begin{subequations}
\be\label{eq.limits}
{\bf r}_1\cdot {\bf r}_2=\cos\theta\,,\quad \theta\in[0,\Delta]\,,\quad \alpha\in[-\delta,\delta)\,,\,\quad|\delta|,|\Delta|\leq\pi\,,
\ee
where $(\Delta,\delta)$ limit the uncertainty in the correlation between the systems, and $\alpha$ fills the same function as in figure \ref{f.dofs}. Without loss of generality, all other entities of \eqref{eq.generalO} can be defined relative to the basis elements of the first system,
\begin{align}
&\hat e_{1,i}=\hat e_i\,,\quad\hat e_{2,i}=R_u(\theta)R_{\hat x}(\alpha)\hat e_i\,,\quad {\bf u}=(0,-\sin\phi_u,\cos\phi_u)\,,\label{eq.rotex}\\
&{\bf a}=R_v(\tau)\hat x\,, \quad {\bf b}=R_v(\tau)R_{\hat x}(\beta)R_{\hat z}(\theta_{ab})\hat x\,,\quad {\bf v}=(0,-\sin\phi_v,\cos\phi_v)\,,\\
& \beta,\phi_u,\phi_v\in[0,2\pi)\,,\quad\tau\in[0,\pi]\,,\quad {\bf a}\cdot {\bf b}=\cos(\theta_{ab})\,.
\end{align}
\end{subequations}
where $(\hat e_1,\hat e_2,\hat e_3)\equiv(\hat x,\hat y,\hat z)$ and ${\bf r}_1=\hat x$. The rotation matrix $R_u(\theta)$ describes a rotation around ${\bf u}$ with an angle of $\theta$. The different vectors $({\bf u, v})$ and angles $(\tau,\beta)$ describe the different possible positions of ${\bf a,b}$ and $\{\hat e_{2,i}\}$ relative to the basis elements of the first system, within the range allowed by \eqref{eq.limits}. For example, the basis elements of the second system are obtained from those of the first by a rotation by $\alpha$ around $\hat x$, followed by a rotation by $\theta$ around ${\bf u}$, as specified in \eqref{eq.rotex}.

The evaluation of \eqref{eq.generalO} contains an averaging over the $({\bf a,b})$ positions by $(\tau,\beta,\phi_v)$, separately from the relative spin system positions, and is non-zero only for combinations $a_ib_j$ with $i=j$. These average to $\cos\theta_{ab}/2$ for $i=x$ and to $\cos\theta_{ab}/4$ for $i=y,z$. Consequently, only the diagonal of $m=R_u(\theta)R_{\hat x}(\alpha)$ contributes to the overlap in \eqref{eq.generalO}, dependent on $(\theta,\alpha,\phi_u)$ through
\be
m_{xx}=\cos\theta\,,\quad m_{yy}+m_{zz}=(1+\cos\theta)\cos\alpha\,,\quad m_{zz}-m_{yy}=(1-\cos\theta)\cos(2\phi_u-\alpha)\,.
\ee
The variable $\theta$ gives the angle deviation between the two systems, $\phi_u$ sets the rotation vector relative to the first system, and $\alpha$ describes the relative rotation of the $(\hat y,\hat z)$ basis. 

The overlap is maximal for $\alpha\equiv0$. A first approximation of a more realistic scenario where increased correlation is obtained through interaction (by a successively increased overlap between the two states) is given by $\theta\in[0,\Delta]$ and $\alpha\in[-\delta,\delta]$, with uncorrelated uncertainties in the separate variables and constant probability distributions in the considered intervals of $(\theta,\alpha)$. The overlap then is
\be\label{eq.avT}
\left\langle S_1({\bf a})S_2({\bf b})\right\rangle=\langle s_1s_2\rangle\left(\frac{\sin\delta}{4\delta}+\frac{\sin\Delta}{2\Delta}+\frac{\sin\Delta\sin\delta}{4\Delta\delta}\right)({\bf a\cdot b})\,.
\ee
In the limit where $\delta\rightarrow0$ and $s_2\rightarrow -s_1$, this becomes $(-{\bf a\cdot b})(\Delta+3\sin\Delta)/(4\Delta)$. Without one-to-one correlations for $s$ and $\alpha\equiv0$, the absolute value of the factor in front of ${\bf a\cdot b}$ is smaller\footnote{Disregarding possible effects of correlated variables with non-constant probability distributions.}. For example, for a random $\alpha$ (i.e. $\delta=\pi$) it is $\sin\Delta/(2\Delta)$, by that only $a_xb_xm_{xx}$ contributes. This last example does not encompass entanglement, but \eqref{eq.2avT} and \eqref{eq.avT} do. These are examples of in which sense entanglement would have a tolerance for deviations away from pair production correlations in the orthogonal information model, which also constitutes a more realistic scenario for general entangling/interacting processes than \eqref{eq.3co}.

\providecommand{\href}[2]{#2}\begingroup\raggedright\endgroup


\begin{thebibliography}{10}

\bibitem{Karlsson:2019vpr}
A.~Karlsson, {{Local, non-classical model of Bell correlations}}, 2019,
  [\href{http://arxiv.org/abs/arXiv:1907.11805}{{arXiv:1907.11805
  [quant-ph]}}].

\bibitem{PhysRevA.61.052306}
V.~Coffman, J.~Kundu and W.~K. Wootters, {Distributed entanglement},
  \href{http://dx.doi.org/10.1103/PhysRevA.61.052306}{Phys. Rev. A {\bf 61},
  052306, 2000},
  [\href{http://arxiv.org/abs/arXiv:quant-ph/9907047}{{arXiv:quant-ph/9907047}}].

\bibitem{PhysRevLett.117.060501}
C.~Lancien, S.~Di~Martino, M.~Huber, M.~Piani, G.~Adesso and A.~Winter, {Should
  entanglement measures be monogamous or faithful?},
  \href{http://dx.doi.org/10.1103/PhysRevLett.117.060501}{Phys. Rev. Lett. {\bf
  117}, 060501, 2016}.

\bibitem{Marletto:2019}
C.~Marletto, V.~Vedral, S.~Virz\`i, E.~Rebufello, A.~Avella, F.~Piacentini,
  M.~Gramegna, I.~P. Degiovanni and M.~Genovese, {{Theoretical description and
  experimental simulation of quantum entanglement near open time-like curves
  via pseudo-density operators}},
  \href{http://dx.doi.org/10.1038/s41467-018-08100-1}{Nat. Commun. {\bf 10},
  182, 2019}.

\bibitem{Bennett:1996gf}
C.~H. Bennett, D.~P. DiVincenzo, J.~A. Smolin and W.~K. Wootters, {{Mixed state
  entanglement and quantum error correction}},
  \href{http://dx.doi.org/10.1103/PhysRevA.54.3824}{Phys. Rev. {\bf A54},
  3824--3851, 1996},
  [\href{http://arxiv.org/abs/arXiv:quant-ph/9604024}{{arXiv:quant-ph/9604024
  [quant-ph]}}].

\bibitem{Bell:1964kc}
J.~S. Bell, {{On the Einstein Podolsky Rosen paradox}},
  \href{http://dx.doi.org/10.1103/PhysicsPhysiqueFizika.1.195}{Physics Physique
  Fizika {\bf 1}, 195, 1964}.

\bibitem{Page:1993wv}
D.~N. Page, {{Information in black hole radiation}},
  \href{http://dx.doi.org/10.1103/PhysRevLett.71.3743}{Phys. Rev. Lett. {\bf
  71}, 3743, 1993},
  [\href{http://arxiv.org/abs/arXiv:hep-th/9306083}{{arXiv:hep-th/9306083
  [hep-th]}}].

\bibitem{Page:2013dx}
D.~N. Page, {{Time Dependence of Hawking Radiation Entropy}},
  \href{http://dx.doi.org/10.1088/1475-7516/2013/09/028}{JCAP {\bf 1309}, 028,
  2013}, [\href{http://arxiv.org/abs/arXiv:1301.4995}{{arXiv:1301.4995
  [hep-th]}}].

\bibitem{Penington:2019npb}
G.~Penington, {{Entanglement Wedge Reconstruction and the Information
  Paradox}},  2019,
  [\href{http://arxiv.org/abs/arXiv:1905.08255}{{arXiv:1905.08255 [hep-th]}}].
  
\bibitem{Almheiri:2019psf}
A.~Almheiri, N.~Engelhardt, D.~Marolf and H.~Maxfield, {{The entropy of bulk
  quantum fields and the entanglement wedge of an evaporating black hole}},
  \href{http://dx.doi.org/10.1007/JHEP12(2019)063}{JHEP {\bf 12}, 063, 2019},
  [\href{http://arxiv.org/abs/arXiv:1905.08762}{{arXiv:1905.08762 [hep-th]}}].

\bibitem{Almheiri:2019hni}
A.~Almheiri, R.~Mahajan, J.~Maldacena and Y.~Zhao, {{The Page curve of Hawking
  radiation from semiclassical geometry}},  2019,
  [\href{http://arxiv.org/abs/arXiv:1908.10996}{{arXiv:1908.10996 [hep-th]}}].

\bibitem{Einstein:1935rr}
A.~Einstein, B.~Podolsky and N.~Rosen, {{Can Quantum-Mechanical Description of
  Physical Reality Be Considered Complete?}},
  \href{http://dx.doi.org/10.1103/PhysRev.47.777}{Phys. Rev. {\bf 47}, 777,
  1935}.

\bibitem{Einstein:1935tc}
A.~Einstein and N.~Rosen, {{The Particle Problem in the General Theory of
  Relativity}}, \href{http://dx.doi.org/10.1103/PhysRev.48.73}{Phys. Rev. {\bf
  48}, 73, 1935}.

\bibitem{Maldacena:2001kr}
J.~M. Maldacena, {{Eternal black holes in anti-de Sitter}},
  \href{http://dx.doi.org/10.1088/1126-6708/2003/04/021}{JHEP {\bf 04}, 021,
  2003},
  [\href{http://arxiv.org/abs/arXiv:hep-th/0106112}{{arXiv:hep-th/0106112
  [hep-th]}}].

\bibitem{Swingle:2009bg}
B.~Swingle, {{Entanglement Renormalization and Holography}},
  \href{http://dx.doi.org/10.1103/PhysRevD.86.065007}{Phys. Rev. {\bf D86},
  065007, 2012}, [\href{http://arxiv.org/abs/arXiv:0905.1317}{{arXiv:0905.1317
  [cond-mat.str-el]}}].

\bibitem{VanRaamsdonk:2010pw}
M.~Van~Raamsdonk, {{Building up spacetime with quantum entanglement}},
  \href{http://dx.doi.org/10.1007/s10714-010-1034-0,
  10.1142/S0218271810018529}{Gen. Rel. Grav. {\bf 42}, 2323--2329, 2010},
  [\href{http://arxiv.org/abs/arXiv:1005.3035}{{arXiv:1005.3035 [hep-th]}}].

\bibitem{Hartman:2013qma}
T.~Hartman and J.~Maldacena, {{Time Evolution of Entanglement Entropy from
  Black Hole Interiors}}, \href{http://dx.doi.org/10.1007/JHEP05(2013)014}{JHEP
  {\bf 05}, 014, 2013},
  [\href{http://arxiv.org/abs/arXiv:1303.1080}{{arXiv:1303.1080 [hep-th]}}].

\bibitem{Maldacena:2013xja}
J.~Maldacena and L.~Susskind, {{Cool horizons for entangled black holes}},
  \href{http://dx.doi.org/10.1002/prop.201300020}{Fortsch. Phys. {\bf 61}, 781,
  2013}, [\href{http://arxiv.org/abs/arXiv:1306.0533}{{arXiv:1306.0533
  [hep-th]}}].

\bibitem{Akers:2019nfi}
C.~Akers, N.~Engelhardt and D.~Harlow, {{Simple holographic models of black
  hole evaporation}},  2019,
  [\href{http://arxiv.org/abs/arXiv:1910.00972}{{arXiv:1910.00972 [hep-th]}}].

\bibitem{Akers:2019nfy}
C.~Akers and P.~Rath, {{Entanglement Wedge Cross Sections Require Tripartite
  Entanglement}},  2019,
  [\href{http://arxiv.org/abs/arXiv:1911.07852}{{arXiv:1911.07852 [hep-th]}}].

\bibitem{Wootters:1982zz}
W.~K. Wootters and W.~H. Zurek, {{A single quantum cannot be cloned}},
  \href{http://dx.doi.org/10.1038/299802a0}{Nature {\bf 299}, 802, 1982}.

\bibitem{Bombelli:1986rw}
L.~Bombelli, R.~K. Koul, J.~Lee and R.~D. Sorkin, {{Quantum source of entropy
  for black holes}}, \href{http://dx.doi.org/10.1103/PhysRevD.34.373}{Phys.
  Rev. {\bf D34}, 373, 1986}.

\bibitem{Srednicki:1993im}
M.~Srednicki, {{Entropy and area}},
  \href{http://dx.doi.org/10.1103/PhysRevLett.71.666}{Phys. Rev. Lett. {\bf
  71}, 666, 1993},
  [\href{http://arxiv.org/abs/arXiv:hep-th/9303048}{{arXiv:hep-th/9303048}}].

\bibitem{Bekenstein:1973ur}
J.~D. Bekenstein, {{Black holes and entropy}},
  \href{http://dx.doi.org/10.1103/PhysRevD.7.2333}{Phys. Rev. {\bf D7}, 2333,
  1973}.

\bibitem{Hawking:1976de}
S.~W. Hawking, {{Black Holes and Thermodynamics}},
  \href{http://dx.doi.org/10.1103/PhysRevD.13.191}{Phys. Rev. {\bf D13}, 191,
  1976}.

\bibitem{Ryu:2006bv}
S.~Ryu and T.~Takayanagi, {{Holographic derivation of entanglement entropy from
  AdS/CFT}}, \href{http://dx.doi.org/10.1103/PhysRevLett.96.181602}{Phys. Rev.
  Lett. {\bf 96}, 181602, 2006},
  [\href{http://arxiv.org/abs/arXiv:hep-th/0603001}{{arXiv:hep-th/0603001
  [hep-th]}}].

\bibitem{Hubeny:2007xt}
V.~E. Hubeny, M.~Rangamani and T.~Takayanagi, {{A Covariant holographic
  entanglement entropy proposal}},
  \href{http://dx.doi.org/10.1088/1126-6708/2007/07/062}{JHEP {\bf 07}, 062,
  2007}, [\href{http://arxiv.org/abs/arXiv:0705.0016}{{arXiv:0705.0016
  [hep-th]}}].

\bibitem{GHZ}
D.~M.~Greenberger, M.~A.~Horne and A.~Zeilinger, in {{Bell's Theorem, Quantum Theory, 
  and Conceptions of the Universe}}, edited by M.~Kafatos (Kluwer, Dordrecht, 1989), p.69,
  [\href{http://arxiv.org/abs/arXiv:0712.0921}{{arXiv:0712.0921 [hep-ph]}}].

\bibitem{Karlsson:2018tod}
A.~Karlsson, {{Space-time emergence from individual interactions}},  2018,
  [\href{http://arxiv.org/abs/arXiv:1806.05710}{{arXiv:1806.05710 [hep-th]}}].

\bibitem{Clauser:1969ny}
J.~F. Clauser, M.~A. Horne, A.~Shimony and R.~A. Holt, {{Proposed experiment to
  test local hidden variable theories}},
  \href{http://dx.doi.org/10.1103/PhysRevLett.23.880}{Phys. Rev. Lett. {\bf
  23}, 880, 1969}.

\end{thebibliography}
\end{document}